\documentclass[sigconf,nonacm]{acmart}
\AtBeginDocument{%
  }

\begin{document}

\title{Neural Network-Driven Volatility Drag Mitigation under Aggressive Leverage}

\author{Christian Bongiorno}
\email{christian.bongiorno@centralesupelec.fr}
\affiliation{%
  \institution{Université Paris-Saclay, CentraleSupélec, Mathématiques et Informatique pour la Complexité et les Systèmes, }
  \city{91190, Gif-sur-Yvette}
  \state{}
  \country{France}
}

\author{Efstratios Manolakis}
\affiliation{%
  \institution{Dipartimento di Fisica e Astronomia “Ettore Majorana”}
  \city{Catania}
  \country{Italy}}

\author{Rosario Nunzio Mantegna}
\affiliation{%
  \institution{Dipartimento di Fisica e Chimica}
  \city{Palermo}
  \country{Italy}}
\affiliation{%
  \institution{Complexity Science Hub}
  \city{Vienna}
  \country{Austria}}

\renewcommand{\shortauthors}{Bongiorno et al.}

\begin{abstract}
This paper introduces a compact reformulation of a modular end-to-end neural network for global minimum-variance portfolio optimization that decouples model complexity from both look-back window length and universe size. A five-parameter hyperbolic weighted moving average combined with a saturating exponential replaces the original 2,400-parameter lag-transformation layer, and a bidirectional gated-recurrent-unit eigencleaning module together with a streamlined marginal-volatility network reduce total learnable parameters from 39,586 to just 2,175. In out-of-sample tests against state-of-the-art nonlinear-shrinkage and risk-parity benchmarks, the compact network attains the lowest realized portfolio variance without compromising expected return. Under long-only constraints, the variance reduction supports substantially higher leverage while maintaining comparable drawdown control. Validation in a high-fidelity trading simulator that incorporates realistic margin-call dynamics confirms enhanced over-leverage resilience. These findings demonstrate that end-to-end variance-minimization architectures can achieve substantial parameter efficiency and robust capital-efficiency gains without sacrificing risk-adjusted performance.

\end{abstract}

\keywords{Neural Network, Nonlinear Shrinkage, Global Minimum Variance, Leverage}


\maketitle

\section{Introduction}

In modern quantitative finance, portfolio optimization remains a cornerstone for balancing the trade‐off between expected return and risk \cite{Rubinstein2002}. By systematically allocating capital across assets, investors seek to construct portfolios that achieve a desired level of return for a given degree of uncertainty. Since Markowitz’s foundational mean–variance framework, numerous extensions have targeted improvements in estimation accuracy, model robustness, and practical implementability.

A fundamental challenge in mean–variance optimization lies in accurately estimating both asset returns and the covariance matrix \cite{Garlappi2007}. Whereas volatility can often be measured with relatively high precision from high‐frequency price data, expected returns exhibit low signal‐to‐noise ratios and are highly non‐stationary. Static estimation techniques, which treat returns as a fixed‐parameter estimation problem, fail to adapt to evolving market regimes and may lead to suboptimal allocations \cite{Chopra1993}.

Neural network models offer a promising avenue for addressing non‐stationarity by jointly learning estimation and forecasting tasks \cite{lee2024overview}. Unlike traditional approaches, such as non‐linear least‐squares methods for covariance estimation, an end‐to‐end neural architecture can ingest raw price or return series and output optimized weights that implicitly capture time‐varying dynamics. In doing so, the model shifts from treating prediction and optimization as separate steps to a unified learning objective that directly minimizes realized portfolio variance.

Lowering realized portfolio volatility has a twofold benefit: it boosts risk‐adjusted performance measures (e.g., the Sharpe ratio) and enhances robustness under leverage. Empirical studies have shown that minimum‐variance portfolios often outperform alternative rules in both volatility reduction and cumulative return, owing to their implicit structural regularization of asset weights \cite{Clarke2006}.

When leverage is introduced, volatility drag—the erosion of compounded returns due to fluctuating portfolio value—becomes a critical consideration. High‐leverage strategies amplify both gains and losses, such that portfolios with elevated variance can suffer disproportionately from drawdowns \cite{JacobsLevy2014}. Thus, reducing baseline volatility is essential for enabling safe leverage expansion without incurring prohibitive drawdowns.

A further practical constraint arises from margin requirements: leveraged positions may trigger margin calls under adverse price movements, forcing deleveraging at inopportune times. Any optimization method intended for high‐leverage applications must therefore account for worst‐case price scenarios and ensure that margin buffers remain adequate throughout the investment horizon.

Bongiorno et al.~\cite{Bongiorno2025} introduced an end‐to‐end neural network for global minimum‐variance optimization that integrated a parameterized lag‐transformation layer, an eigencleaning module, and a marginal‐volatility network. While effective in reducing realized variance, that model employed approximately $40,000$ learnable parameters and was calibrated for a fixed in‐sample window length, limiting its scalability and adaptability to different market universes.

In this paper, we propose a compact reformulation that reduces the total number of learnable parameters to approximately $2,000$ and decouples model complexity from both the number of look‐back days and the size of the asset universe. By replacing the high‐dimensional lag‐transformation weights with a five‐parameter formulation based on hyperbolic weighted moving averages and a saturating exponential, we achieve dramatic efficiency gains while preserving variance‐minimization performance.

Finally, we validate our approach within a high‐fidelity operational simulator that executes market orders at the close price using the most recent available information. The simulator incorporates conservative worst‐case price scenarios to model margin‐call dynamics, thereby providing a realistic assessment of over‐leverage resilience up to 4:1. Through these experiments, we demonstrate that our parameter‐efficient network enables higher leverage deployment without sacrificing drawdown control.

\section{Volatility Drag under Leverage}\label{sec:vol_drag}
In this work, we consider a rebalanced leveraged portfolio in which, at the close of each trading day, holdings are adjusted back to a fixed weight vector $\mathbf{w}$ and the overall leverage factor $\ell>0$ is reinstated via borrowing or cash adjustments.

Denoting by $\mathbf r_t$ the vector of (unlevered) asset returns over day $t$, the first two moments of the portfolio returns are

\begin{equation}
 \mathbb{E}[\mathbf{w}^{\top}\mathbf{r}_{t}] = \mu,
\qquad 
\mathbb{V}[\mathbf{w}^{\top}\mathbf{r}_{t}] =  \mathbf{w}^{\top}\boldsymbol{\Sigma}\mathbf{w}=\sigma^{2}, 
\end{equation}

where $\boldsymbol{\Sigma}$ is the population covariance matrix.
Introducing a constant leverage factor $\ell$, assuming a zero risk-free rate, the compounded portfolio equity (net asset value) after $\Delta t$ periods is

\begin{equation}
    m_{\Delta t}
 = m_{0}\prod_{t=1}^{\Delta t}\left( 1 + \ell\, \mathbf{w}^\top \mathbf{r}_t\right)
 = m_{0}\exp\! \left[ \sum_{t=1}^{\Delta t}\ln \!\left(1+\ell\,   \mathbf{w}^\top \mathbf{r}_t \right)\right].
\end{equation}

When returns are small relative to 1, a second-order expansion $\ln(1+x)\simeq x-\tfrac{1}{2}x^{2}$ yields

\begin{equation}\label{eq:growth_rate}
    \mathbb{E}\!\bigl[\ln(1+\ell\, \mathbf{w}^\top \mathbf{r}_t )\bigr]
  \;\simeq\;
  \ell\,\mu - \tfrac{1}{2}\,\ell^{2}\sigma^{2}.
\end{equation}

The second term in Eq.~\eqref{eq:growth_rate} is the volatility drag, i.e., the reduction in compound growth caused by return variability.  Practically, volatility drag scales as $\tfrac{1}{2}\ell^{2}\sigma^{2}$, so doubling leverage quadruples the drag.  Minimising $\sigma^{2}$ is therefore a prerequisite for sustainable high leverage.

\section{Model Architecture}

The end‐to‐end neural network for global minimum‐variance optimization consists of four sequential stages. First, raw historical returns for \(n\) assets over a look‐back window of length \(\Delta t_\textrm{in}\) are fed into a parameterized lag‐transformation block, which produces a temporally enriched feature tensor (Sec.~\ref{sec:lagtrans}). Second, this tensor branches into two parallel modules: a correlation‐denoising network that outputs a cleaned inverse eigenvalue spectrum (Sec.~\ref{sec:corrclean}) and a marginal‐volatility network that estimates inverse asset‐specific volatilities (Sec.~\ref{sec:volatility}). Third, the outputs of these branches are recombined to assemble an approximate inverse covariance matrix
\begin{equation}\label{eq:NNcov}
\mathbf{\Sigma}^{-1}_\mathrm{NN}=\mathbf{D}^{-1}_\mathrm{NN} \mathbf{C}^{-1}_\mathrm{NN} \mathbf{D}^{-1}_\mathrm{NN}.
\end{equation}

Fourth, we compute the analytic global minimum‐variance weights
\begin{equation}\label{eq:GMV}
\mathbf{w} \;=\;\frac{\mathbf{\Sigma}^{-1}_\mathrm{NN}\,\mathbf{1}}{\mathbf{1}^\top \mathbf{\Sigma}^{-1}_\mathrm{NN}\,\mathbf{1}},
\end{equation}
where \(\mathbf{1}\) is the \(n\)-vector of ones. Importantly, by using fixed‐size parameterizations in each module, the total number of learnable parameters is independent of both the temporal depth \(\Delta t_\mathrm{in}\) and the cross‐sectional dimension \(n\), enabling immediate transfer to new asset universes or sampling frequencies without retraining.

We train the entire network by minimizing the realized variance of its portfolio returns. Denote by \(\mathbf{\Sigma}_\mathrm{out}\in\mathbb{R}^{n \times n}\) the future realized covariance matrix of asset returns $\mathbf{R}_\mathrm{out}\in \mathbb{R}^{\Delta t_\mathrm{out} \times n}$ . The loss function is defined as
\begin{equation}\label{eq:loss}
\mathcal{L}( \mathbf{w}, \mathbf{\Sigma}_\mathrm{out}) \;=\; n \, \mathbf{w}^\top \mathbf{\Sigma}_\mathrm{out} \mathbf{w},
\end{equation}
which corresponds to the out-of-sample variance of the model’s portfolio returns. No explicit expected‐return term is included, allowing the network to focus solely on variance reduction. Detailed parameterizations and learning dynamics for each module are described in the following subsections. 

It is worth stressing that the trained covariance $\boldsymbol{\Sigma}_\textrm{NN}$ can be extracted from the calibrated model and tested with constrained optimizations, such as in this work, where we apply it to the long-only portfolios.

\subsection{Lag-Transformation Module}\label{sec:lagtrans}

Let the input return series be denoted by 
$\mathbf{R} \in \mathbb{R}^{\Delta t_\mathrm{in} \times n}$,
where each $\mathbf{r}_{t}\in\mathbb{R}^n$ is the vector of asset returns at lag $t$, with $t=1$ corresponding to the most recent observation. The lag‐transformation module produces an output matrix 
$\widetilde{\mathbf{R}} \in \mathbb{R}^{\Delta t_\mathrm{in} \times n}$
by applying a parametric temporal weighting and nonlinear clipping to each lag. Crucially, this module is parameterized by only five scalars 
\(\boldsymbol{\theta} = \{\theta_{1}, \theta_{2}, \theta_3, \theta_{4}, \theta_{5}\} \in \mathbb{R}_{>0}^5\) , making its complexity independent of both $\Delta t_\mathrm{in}$ and $n$. These five scalars are treated as model parameters and are learned jointly with the rest of the network via backpropagation and gradient descent optimization.

For each lag $t\in\{1,\dots,\Delta t_\mathrm{in}\}$, define the hyperbolic weight
\begin{equation}\label{eq:alpha}
\alpha_{t} \;=\; \theta_{1}\,t^{-\theta_{2}},    
\end{equation}

and the saturating exponential threshold
\begin{equation}\label{eq:beta}
\beta_{t} \;=\; \theta_{3}\;-\;\theta_{4}\,e^{-\,\theta_{5}\,t}\!,    
\end{equation}

where all operations are applied elementwise across assets. The transformed feature for asset $i$ at lag $t$ is given by
\begin{equation}\label{eq:lagtransform}
\widetilde{r}_{t,i}
\;=\;
\frac{\alpha_{t}}{\beta_{t} }
\,\tanh \bigl(\beta_{t}\,r_{t,i}\bigr) \quad \text{with} \quad i=1, \dots, n.    
\end{equation}

This formulation captures the empirically observed hyperbolic decay and exponential saturation behaviors \cite{Bongiorno2025} while requiring only five learned parameters, and thus scales trivially to different window lengths and asset universes.

\subsection{Correlation Cleaning Module}\label{sec:corrclean}
We denote by \(\mathbf{C}\in\mathbb{R}^{n\times n}\) the empirical correlation matrix computed from the lag‐transformed features of Eq.~\eqref{eq:lagtransform}. Writing its eigendecomposition as
\begin{equation}\label{eq:spectral}    
\mathbf{C} \;=\; \mathbf{Q}\,\boldsymbol{\Lambda}\,\mathbf{Q}^\top,
\qquad
\boldsymbol{\Lambda} \;=\;\mathrm{Diag}\bigl(\lambda_{1},\dots,\lambda_{n}\bigr),
\end{equation}
we seek a rotation‐invariant, permutationally equivariant mapping
$\boldsymbol{\Lambda} \mapsto \boldsymbol{\Lambda}_\textrm{NN}^{-1}$ that produces cleaned inverse eigenvalues \cite{Bongiorno2025}. In our compact model, this mapping is still implemented by a 
bidirectional Recurrent Neural Network (RNN).

Interpreting the ordered spectrum as a sequence, we embed each eigenvalue at rank $i$ into a unified feature vector  
\begin{equation}
\mathbf{x}_i \;=\;\left\{\lambda_i,\;q,\;\sqrt{n},\;\sqrt{\Delta t_{\mathrm{in}}}\right\} \quad \text{with} \quad i=1, \dots, n.    
\end{equation}
The first component conveys the raw eigenvalue, while the remaining entries respectively encode the aspect ratio $q=n/\Delta t_{\mathrm{in}}$ and the two natural scales $\sqrt{n}$ and $\sqrt{\Delta t_{\mathrm{in}}}$, which together capture both the matrix shape and the magnitude of sampling noise. Although $n/\Delta t_{\mathrm{in}}$ it is algebraically redundant given  $\sqrt{n}$ and $\sqrt{\Delta t_{\mathrm{in}}}$, its explicit inclusion empirically accelerates convergence during training.  

Then, we process the sequence $\{\mathbf{x}_1,\dots,\mathbf{x}_n\}$ with a Bidirectional Gated Recurrent Unit (BiGRU) network having $k=16$ hidden units per direction.  Specifically, for $i=1,\dots,n$ we compute  
\begin{equation}\label{eq:gru}
\boldsymbol{h}_i^{\rightarrow}
\;=\;
\mathrm{GRU}_{k}^{\rightarrow}\bigl(\boldsymbol{h}_{i-1}^{\rightarrow},\, \mathbf{x}_i\bigr),
\qquad
\boldsymbol{h}_i^{\leftarrow}
\;=\;
\mathrm{GRU}_{k}^{\leftarrow}\bigl(\boldsymbol{h}_{i+1}^{\leftarrow},\,  \boldsymbol{x}_i \bigr),
\end{equation}
 which we concatenate as \(\boldsymbol{h}_i = [\boldsymbol{h}_i^{\rightarrow};\,\boldsymbol{h}_i^{\leftarrow}]\). A shared affine projection followed by a softplus activation then yields
\begin{equation}
\lambda_{i,\textrm{NN}}^{-1}
\;=\;
\mathrm{softplus}\bigl(\boldsymbol{\gamma}^\top \boldsymbol{h}_i + \omega \bigr), \quad \text{with}\quad i=1, \dots, n.
\end{equation}
The cleaned inverse correlation is reconstructed as
\begin{equation}\label{eq:cleanedCorr}
\mathbf{C}_{\mathrm{NN}}^{-1}
\;=\;
\mathbf{Q}\,\mathrm{Diag}\bigl(\lambda_{1,\textrm{NN}}^{-1},\dots,\lambda_{n,\textrm{NN}}^{-1}\bigr)\,\mathbf{Q}^\top.
\end{equation}

By replacing the original bidirectional LSTM of Ref.~\cite{Bongiorno2025} with a GRU of $k=16$ units per direction, we reduce the parameter count of this module from over 30 000 to under 2 000, without degrading expressivity.

Although classical NonLinear Shrinkage (NLS) theory prescribes an analytical form for the optimal shrinkage in the high‐dimensional limit, our BiGRU learns an implicit shrinkage function end‐to‐end under the realized‐variance loss. This architecture has been shown to be highly effective in Ref.~\cite{Bongiorno2025}, where a BiRNN‐based spectral denoiser outperformed standard NLS estimators in high‐dimensional covariance cleaning.

\subsection{Marginal Volatility Module}\label{sec:volatility}
In our architecture, the marginal‐volatility module converts each asset’s sample standard deviation $\left\{ \widetilde{\sigma}_1,\dots, \widetilde{\sigma}_n\right\}$ computed over the lag‐transformed return series $\widetilde{\mathbf{R}}\in\mathbb{R}^{\Delta t_\mathrm{in}\times n}$ of Eq.~\eqref{eq:lagtransform} into an inverse volatility scale by means of a small, per‐asset neural network. Concretely, for each asset $i$ we feed its empirical standard $\widetilde{\sigma}_i$ deviation into a MultiLayer Perceptron (MLP) with a single hidden layer of $8$ neurons using a leaky ReLU activation, followed by a softplus output to enforce positivity. Because this network is applied independently to each asset, no cross‐asset information is exchanged in this step. The resulting vector of inverse volatilities
\begin{equation}
    \mathbf{D}^{-1}_\textrm{NN}=\text{Diag}\left(\sigma_{1,\textrm{NN}}^{-1}, \dots, \sigma_{n,\textrm{NN}}^{-1} \right)
\end{equation}

\section{Real‑World Experimental Setup}
The experimental framework that follows is designed for direct deployment in production environments. Sec~\ref{sec:stock} presents the universe-selection protocol, which applies market-capitalization and liquidity filters at each rebalancing date and relies exclusively on information available at selection time to eliminate any look-ahead bias. Sec.~\ref {sec:train} outlines the training pipeline on historical data crafted to ensure robust out-of-sample generalization of the neural network. Sec.~\ref{sec:simulator} describes the high-fidelity simulator that reproduces the complete lifecycle of an Interactive Brokers cash-and-margin account. Finally, Sec.~\ref{sec:others} specifies the alternative benchmark portfolios employed in this study.

\subsection{Stock Selection}\label{sec:stock}
We define our investable universe to comprise all U.S. common equities and ADRs listed on the NYSE or NASDAQ between January 1, 1990 and December 31, 2024, explicitly excluding funds and ETFs. To eliminate look-ahead bias, at each rebalance date \(t\) we use only information available up to \(t-1\) and drop any security with a delisting within the next \(\Delta t_{\rm out}=5\) trading days. From this broad set, we apply a two-tiered filter that enforces (i) homogeneous long-term trading records over our five-year calibration window (\(\Delta t_{\rm in}=1200\) days) and (ii) robust short-term liquidity over the most recent 5–20 trading days.

In the first tier, we require that in every rolling one-year subperiod of the calibration window, a security participates in at least 95\% of closing auctions, ensuring continuous price histories. In the second tier, designed to capture extreme liquidity shortfalls immediately prior to trading, we mandate that over the past five days (a) 100\% of closing auctions execute, (b) average daily volume represents at least 1\% of shares outstanding, and (c) average daily notional turnover represents at least 1\% of market capitalization. We further exclude any stock whose shares outstanding fall below 5 million or whose closing price lies outside the \$10–\$2 000 range one day before each rebalance.

To guard against univariate outliers and redundant listings, we then remove all securities whose log-standard deviation of returns falls below the 1.5 Inter-Quartile Range (IQR) threshold, computed separately over the most recent 5-day and 20-day windows; thus avoiding assets with anomalously low risk. This filter prevents the optimization from degenerating into a trivial univariate variance minimization and preserves the truly multivariate nature of our method, ensuring that training and testing leverage cross-asset interactions rather than isolated low-risk outliers. Eventually, we collapse multiple share classes by retaining only the class with the highest market capitalization as of the prior close, and we eliminate one member of any pair whose in-sample return correlation exceeds 0.95 \cite{engle2019large}. Finally, we rank the remaining candidates by previous-day market capitalization and select the top \(n=1000\) to form the daily rebalancing universe.

\subsection{Training Procedure}\label{sec:train}
The training protocol largely follows that of Ref.~\cite{Bongiorno2025}, with the sole modification that the in-sample window length $\Delta t_{\mathrm{in}}$ is sampled uniformly from the integer interval $\Delta t_{\mathrm{in}}\in [250,1200]$ for each batch. For each training batch, we uniformly sample the cross-sectional dimension $n\in[50,350]$ and the in-sample window length $\Delta t_{\mathrm{in}}$ to promote model robustness across aspect ratios $q=n/\Delta t_{\mathrm{in}}$.

Given a randomly drawn reference date $t$ such that the interval $[t-\Delta t_{\mathrm{in}}-1,\,t+\Delta t_{\mathrm{out}}]$ is contained within the calibration window for the validation year, we select $n$ assets from the filtered investable universe available at time $t-1$. The input to the model is the matrix of adjusted close-to-close returns over the period $[t-\Delta t_{\mathrm{in}},\,t-1]$. We compute the realized out-of-sample covariance matrix $\boldsymbol{\Sigma}_{\mathrm{out}}$ from returns over the subsequent window $[t,\,t+\Delta t_{\mathrm{out}}]$, where $\Delta t_{\mathrm{out}}=5$. A one-day shift between in- and out-of-sample windows prevents data leakage from end-of-day prices.

We optimize the portfolio weight $\mathbf{w}$ by minimizing the end-to-end loss of Eq.~\eqref{eq:loss}
using the Adam optimizer with an initial learning rate of $10^{-4}$ and an exponential decay factor of $0.99^{1/500}$ per batch. We employ gradient-norm clipping at a threshold of 1.0 to enhance numerical stability. Training proceeds for 100 epochs, each consisting of 500 gradient steps with a batch size of 32. We repeat this procedure for each of the 24 yearly validation windows (2000-2024), yielding a total of 24 trained networks.

\subsection{High-Fidelity Simulator}\label{sec:simulator}
We simulate an Interactive Brokers cash-and-margin account in continuous calendar time, aligning market events (price moves, dividends, corporate actions) to the exchange calendar supplied with our price data. At any date \(t\), the account holds a vector of shares \(s_{t-1} = \{s_{t-1,i}\}_{i=1}^n\) and a cash balance \(c_{t-1}\). Overnight and intraday cash dividends are credited to \(c_t\). If \(c_t\) ever becomes negative, the account incurs debit interest at a daily rate equal to the prevailing Fed Funds effective rate \cite{FederalReserve2025fedfunds} plus the broker’s spread, using a 360-day convention \cite{interactivebrokers_rate}. Positive cash balances above the unpaid \$100 k tranche accrue credit interest according to the broker’s tiered schedule \cite{interactivebrokers_commissions}.

Each rebalance date, before submitting trades, we compute a pre-trade estimate of Net Liquidation Value (NLV)
\begin{equation}\label{eq:expectedNLV}
\widehat{\mathrm{NLV}}'_t = c_{t-1} + \sum_{i=1}^n s_{t-1,i}\,\hat p_{t,i},
\end{equation}
where \(\hat p_{t,i}\) is the primary-exchange opening price on day \(t\) (or, if unavailable, the prior adjusted close). Target shares are then set by
\begin{equation}\label{eq:targetShares}
s_{t,i} = \mathrm{round}\!\left(\ell \, w_{t,i}\,\widehat{\mathrm{NLV}}'_t / \hat p_{t,i}\right) \quad \text{with} \quad w_{t,i} \geq 0,
\end{equation}
and executed as end-of-day market orders at the actual closing price \(p_{t,i}\), yielding the post-trade NLV
\begin{equation}\label{eq:realizedNLV}
    \mathrm{NLV}_t = c_t + \sum_{i=1}^n s_{t,i}\,p_{t,i},
\end{equation}
with the cash balance \(c_t\) adjusted for the day’s dividend cash flows and trading fees.
We emphasize that, in this work, we do not model any market impact: execution is assumed to occur at the displayed prices without affecting the price formation process.

The portfolio weights $\mathbf{w}_t$ are obtained by performing an external Quadratic Programming (QP) optimization with long-only constraints $w_{t,i}\geq 0$ from the NN internal representation of the covariance matrix $\boldsymbol{\Sigma}_\mathrm{NN}$ by inverting Eq.~\eqref{eq:NNcov}. It is important to note that a long-short strategy with a realistic fee structure, although interesting for academic purposes, is less relevant for a practical investing strategy when a reliable guess of the expected returns is not provided. 

Trades incur three categories of costs: per-share commissions following the broker’s tiered schedule (0.035 ¢/share below 300 k shares traded in the month, 0.020 ¢ above, with a 0.35 \$ minimum per ticket); exchange, clearing, and regulatory fees totalling 0.0845 \% of executed notional; and Section 31 SEC fees on sells at 1.157 bp of sell notional \cite{interactivebrokers_commissions}. All fees and interest charges are debited immediately from the cash.

Because we permit up to 4:1 gross leverage (maintenance margin requirement of 25 \%), we enforce an intraday margin‐maintenance check on each trading day. For this purpose, we compute the maintenance margin ratio using each position’s intraday low price, thereby adopting a conservative worst‐case valuation that assumes all stocks simultaneously realize their intraday minima. To avoid overly pessimistic valuations driven by extreme intraday swings (e.g., during the 2010 flash crash \cite{cftc_sec_flashcrash2010}), we floor the low price at

\begin{equation}
\max\bigl(\mathrm{low},\,0.85\min(\mathrm{open},\mathrm{close})\bigr),
\end{equation}
so that no more than a 15 \% loss relative to the open or close is recognized. If this intraday‐low valuation would breach the 25 \% maintenance threshold, we would immediately liquidate sufficient shares, proportionally to the portfolio allocation, at those intraday lows to restore the margin ratio to 27 \% (i.e.\ 2 pp above the requirement), incurring the same per‐share commissions and regulatory fees described above. Although somewhat pessimistic, this low‐price–based enforcement mechanism serves as a valuable stress test for the model.

\section{Compared Estimators}\label{sec:others}
We benchmark our compact neural estimator against three broad families of allocation techniques: univariate-based rules, risk-parity approaches, and covariance-filtering methods solved via long-only quadratic programming.  A naive Equally-Weighted (EW) strategy, which assigns identical weight to each asset, is included too as a simple diversification baseline. 

The first, and probably, most popular univariate rule is the Market-Capitalization Weighting (MCW), which allocates in proportion to each asset’s float-adjusted market value \cite{hsu2004cap};  this practice is very common since it is the basis of the composition of many indices and most of the ETFs. The Equal-Risk Budget (ERB) scales each asset by the reciprocal of its estimated variance \cite{leote2012demystifying}, thereby equalizing univariate risk exposures. 

Equal Risk Contribution (ERC) extends the ERB principle by solving for weights that equalize marginal risk contributions under the full covariance structure \cite{maillard2010properties}. Relative to the GMV portfolio, ERC typically yields more evenly distributed weights across assets and, therefore, ensures exposure to a broader set of systematic risk factors; this balanced risk allocation is often valued in practice for its enhanced diversification benefits and greater robustness to market regime shifts. Hierarchical Risk Parity (HRP) further exploits the empirical clustering of assets by constructing a dendrogram from the sample correlation matrix and then recursively allocating capital within each cluster so as to balance cluster-level variances \cite{lopez2016building}, potentially capturing sectoral relationships and reducing estimation noise.

On the covariance‐filtering side, each estimator first produces a filtered covariance matrix $\hat{\boldsymbol{\Sigma}}$, which we then plug into a long‐only QP to obtain the GMV weights under the constraints $w_i\geq0$ and $\sum_i w_i=1$. The Maximum Likelihood Estimator (MLE) of the sample covariance serves as a high‐noise baseline prior to any shrinkage. The state‐of‐the‐art Quadratic‐Inverse Shrinkage (QIS) estimator of Ledoit–Wolf \cite{ledoit2022quadratic} efficiently regularizes the sample eigenvalues according to Random Matrix Theory (RMT), delivering a well‐conditioned $\boldsymbol{\Sigma}_\textrm{QIS}$ estimator that, when fed into the same long‐only QP, yields markedly improved out‐of‐sample risk forecasts. By contrast, the Average Oracle (AO) method \cite{bongiorno2023filtering} abandons time‐varying adjustments entirely: it calibrates a fixed set of eigenvalues over an extended historical window (1990–2000) in a stock‐agnostic, time‐invariant fashion and then applies this identical eigen‐spectrum unchanged to all future covariance estimates. Remarkably, this deceptively simple static correction, when combined with the same long‐only QP solver,  consistently outperforms more sophisticated filtering schemes, achieving higher Sharpe ratios, lower portfolio turnover, and broader effective diversification in long‐only backtests \cite{bongiorno2024covariance,fermanian2024model,Bongiorno2025}.

\section{Results}

We evaluate our model within a rolling–annual calibration framework: at each trading year, the network is retrained on all data up to the previous year, after which portfolio weights are computed every day under long-only constraints and a fixed leverage \(\ell\) is held constant at each daily rebalance.  The simulation starts at 2000-01-01 with an initial cash capital $c_1=1,000,000$\$, which is compatible with a small hedge fund, and no initial investment. Then the simulation is carried on up to 2024-12-31, and the performances are tracked daily.  The following backtesting considers at any point in time $t$ the full investible universe of $n=1,000$ stocks available at $t$ as described in Sec.~\ref{sec:stock}. If on a given day a stock exits from the investable universe, then the position must be closed that day. 
\begin{figure}
    \centering
    \includegraphics[width=0.9\linewidth]{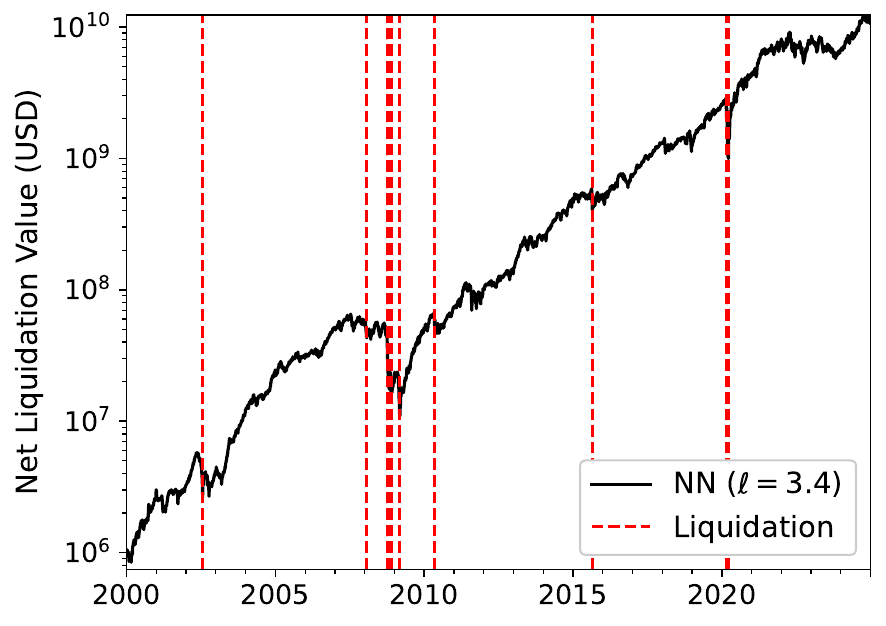}
    \includegraphics[width=0.9\linewidth]{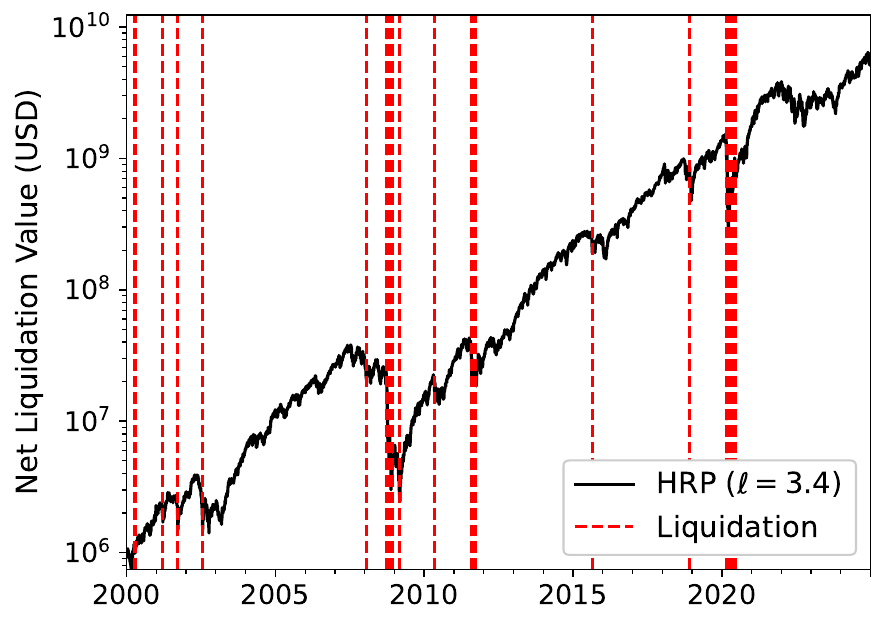}
    \caption{NLV of two high-leverage strategies for NN and HRP on the top 1000 high-capitalized US stocks. The dotted lines represent intraday liquidations. The simulations do not include any market impact.}
    \label{fig:ICAIF-NN-Liquidation}
\end{figure}

Fig.~\ref{fig:ICAIF-NN-Liquidation} compares the NLV of the neural estimator (upper panel) and the HRP benchmark (lower panel) at \(\ell=3.4\). Vertical red dotted lines denote intraday margin‐call‐induced liquidations. Despite operating at high leverage, the neural strategy delays its first forced liquidation and experiences markedly fewer subsequent liquidations than HRP, proving better drawdown control under extreme market stress. It is important to note that our simulator assumes exogenous execution prices and therefore does not model market impact. While such effects are negligible for highly liquid, large‐cap US stocks in closing auctions \cite{salek2023price}, their influence grows with larger NLV. Thus, the full NLV trajectory should be interpreted with caution, while the slope is a more reliable measure when rescaled to lower NLV levels.

\begin{figure}
    \centering
    \includegraphics[width=0.9\linewidth]{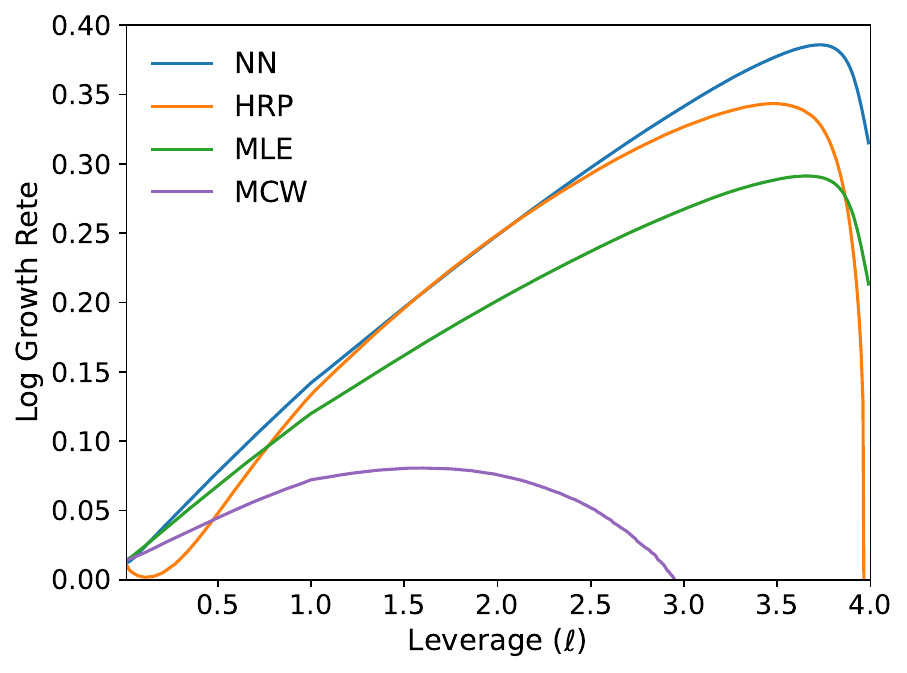}
    \includegraphics[width=0.9\linewidth]{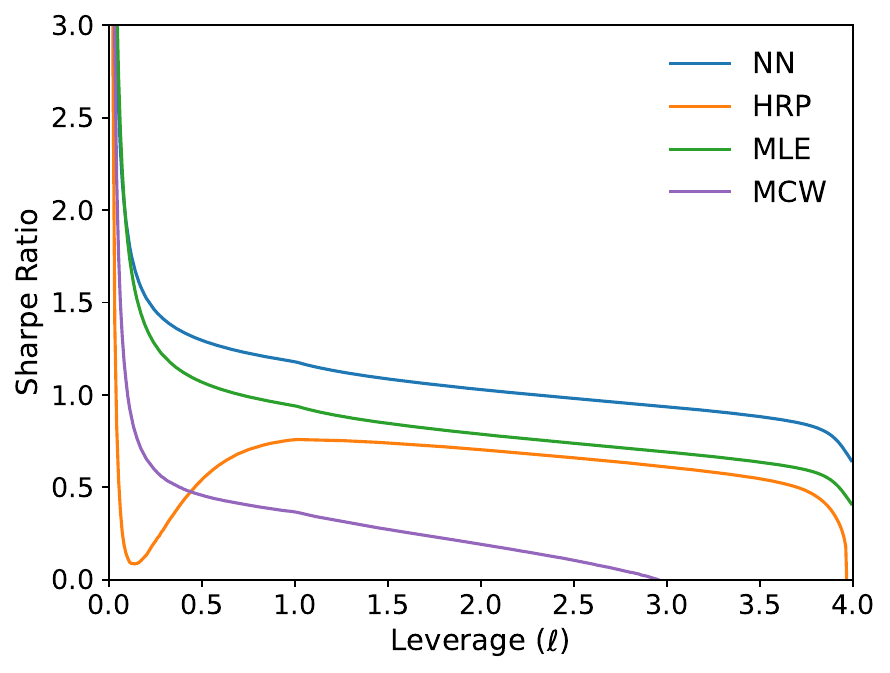}
    \caption{The upper panel represents the out-of-sample log growth rate (see Eq.~\eqref{eq:growth_rate}) measured from the realized NLV of Eq.~\ref{eq:realizedNLV} for different values of the leverage. The lower panel is the Sharpe ratio computed from log returns of the realized NLV as a function of the leverage. Both plots refer to a 25-year simulation of the top 1000 high-capitalized US stocks. }
    \label{fig:ICAIF-LogGrowth-Leverage}
\end{figure}

Fig.~\ref{fig:ICAIF-LogGrowth-Leverage} plot the compound log-growth rate and the Sharpe ratio (computed on log returns) as functions of leverage over a 25-year simulation. In both metrics, the neural estimator consistently outperforms all competing methods, converting additional leverage into incremental return more efficiently while containing volatility drag.

An exhaustive analysis over the full grid of approximately $400$ leverage points ($0.01 \le \ell \le 3.99$ in increments of 0.01) confirms and materially extends the preliminary findings illustrated in Figs.~\ref{fig:ICAIF-LogGrowth-Leverage}–\ref{fig:ICAIF-NN-Liquidation}. Three quantitatively robust regularities emerge by analyzing Tab.~\ref{tab:key_metrics}

First, the neural estimator delays the first forced liquidation to $\ell_{\mathrm{liq}}^{(\mathrm{NN})}=2.77$, whereas all competing covariance filters trigger margin calls between 2.61 and 2.73. Although the gap appears modest, the nonlinearity of liquidation costs and the convex capital‐efficiency penalty beyond $\ell=2.5$ leverage render this extension economically meaningful, corroborating the visual separation between the solid and dashed curves in Figure~\ref{fig:ICAIF-NN-Liquidation}.

Second, within the frictionless region $0.5 \le \ell \le 2.5$, realised volatility grows approximately linearly with leverage, $\sigma(\ell)\simeq a\,\ell$. A least‐squares fit yields the smallest slope for our model, $a_{\mathrm{NN}}=0.1208$, followed by QIS ($0.1284$) and AO ($0.1275$); for comparison, EW requires $0.2089$. Mean return increases with slope $b=\partial\mu/\partial\ell$, and only the neural estimator satisfies $b>a$, yielding an incremental efficiency ratio $b_{\textrm{NN}}/a_{\textrm{NN}}=1.08>1$. All rival methods fall below unity, confirming that the advantage visible in Fig.~\ref{fig:ICAIF-LogGrowth-Leverage} in converting leverage into excess return rather than into incremental risk.

Third, the superiority becomes more pronounced in the long‐only leverage band ($2.5 \le \ell \le 3.5$). At $\ell=3$, the neural strategy realises a Sharpe ratio of $1.12$ with annualised volatility of $0.36$, while the nearest alternative (AO) attains a Sharpe ratio of $0.99$ and $0.38$, respectively. Maximum Draw‐Down (MDD) improves from $-0.83$ to $-0.77$ despite identical liquidation counts. To verify that these differences are not attributable to sampling variability, we construct a Model Confidence Set (MCS) \cite{HansenLundeNason2011MCS} on a Sharpe-based loss: for each method, we compute a daily Sharpe proxy by standardising returns with an exponentially weighted moving-average volatility with memory $0.94$ \cite{RiskMetrics2006}, and form the $5\%$ MCS using a stationary block bootstrap with optimally selected block lengths \cite{Patton30012009} and 100,000 resamples. Under this procedure, only the NN strategy belongs to the superior set at $\ell=3$ (Table \ref{tab:key_metrics}). When the legacy NN \cite{Bongiorno2025} is added to the candidate set, the MCS does not reject equal performance and, therefore, does not distinguish the two architectures.

\begin{table}
\centering

\begin{tabular}{lcccccc}
\toprule
Method & $\ell_{\mathrm{liq}}$ & $b/a$ & Sharpe$_{3}$ & Vol$_{3}$ & MDD$_{3}$  &p-val$_3$\\
\midrule
\textbf{NN}& \textbf{2.77} & \textbf{1.08} & \textbf{1.12} & \textbf{0.36} & $\mathbf{-0.77}$  &$\mathbf{1.000}$\\
AO   & $2.73$ & $0.95$ & $0.99$ & $0.38$ & $-0.83$  &$0.013$\\
HRP  & $2.66$ & $0.94$ & $0.88$ & $0.53$ & $-0.89$  &$0.001$\\
ERC   & $2.63$ & $0.89$ & $0.89$ & $0.57$ & $-0.90$  &$0.002$\\
EW   & $2.62$ & $0.86$ & $0.86$ & $0.63$ & $-0.92$  &$0.001$\\
MLE  & $2.69 $& $0.85$ & $0.89$ & $0.38$ & $-0.86$  &$0.002$\\
QIS  & $2.70$ & $0.82$ & $0.87$ & $0.39$ & $-0.85$  &$0.002$\\
ERB  & $2.63$ & $0.78$ & $0.79$ & $0.55$ & $-0.91$  &$0.000$\\
MCW & $2.61$ & $0.32$ & $0.28$ & $0.59$ & $-0.98$  &$0.000$\\
\bottomrule
\end{tabular}
\caption{Key risk metrics by covariance estimator.  $\ell_{\mathrm{liq}}$ is the lowest leverage at which the first forced liquidation occurs; $b/a$ is the incremental efficiency ratio (slope of mean return divided by slope of volatility) estimated over $0.5\le\ell\le2.5$; Sharpe$_{3}$, Vol$_{3}$, and MDD$_{3}$ are the realised Sharpe ratio, annualised volatility, and maximum draw‐down at $\ell=3$.  The last column reports the MCS inclusion p-value at $\ell=3$, computed on the Sharpe-based loss; larger values indicate stronger evidence that the method belongs to the superior set. Boldface marks the best value in each column; for the MCS column, boldface denotes inclusion superior set at $95\%$ confidence level.}\label{tab:key_metrics}
\end{table}

Taken together, these results confirm that the proposed compact neural estimator not only minimises variance at unit leverage but also extends the safe‐leverage corridor, improves incremental risk‐adjusted efficiency, and contains drawdowns at high leverage, thereby translating its parameter efficiency into genuine capital‐efficiency gains.

\section{Discussions}
In this paper, we have introduced a compact end-to-end neural network architecture for global minimum-variance portfolio optimization that decouples model complexity from both the look-back window length and the size of the asset universe. By replacing a high-dimensional lag-transformation layer with a five-parameter hyperbolic weighted moving average combined with a saturating exponential, and by substituting a bidirectional gated-recurrent-unit eigencleaning module and a streamlined marginal-volatility network for larger, more cumbersome components, we reduce the total number of learnable parameters from nearly 40,000 to approximately 2,000. This parameter efficiency enables immediate transfer to new universes or sampling frequencies without retraining, while preserving the network’s ability to directly minimize realized portfolio variance.

Extensive out-of-sample experiments against state-of-the-art benchmarks—including nonlinear shrinkage estimators and risk-parity rules—demonstrate that our compact network not only attains the lowest realized variance at unit leverage but also extends the safe-leverage corridor, improving the first forced-liquidation threshold from approximately 2.61–2.73 under competing filters to 2.77 under our model. Within the frictionless leverage band, the network achieves an incremental efficiency ratio $b/a>1$, indicating that incremental mean return grows faster than volatility, and at high-leverage $\ell=3$, it delivers a Sharpe ratio in excess of 1.1 with tighter drawdown control. High-fidelity simulator tests that incorporate realistic margin dynamics and transaction costs confirm the network’s enhanced over-leverage resilience up to 4:1 gross leverage. We note, however, that our simulator currently assumes exogenous execution prices and does not model market impact; under large orders or stressed liquidity conditions, this simplification may understate realized costs and drawdown effects.

While the compact architecture is purpose-built for variance minimization on daily data and excels in that setting, the original network remains preferable when the objective, constraints, or application context are expected to change. In particular, if one intends to modify the loss function (e.g., to incorporate return-predictive components or alternative risk measures), operate at different sampling frequencies, or adapt to domain-specific structure, the original model’s higher parameterization and more general lag-transformation afford greater flexibility. We regard the two architectures as complementary: the compact model is an efficient, task-optimized variance minimizer for daily horizons, whereas the original network provides a more elastic platform for extending the objective or porting to nonstandard environments.

Future work could address extending the portfolio-variance loss to incorporate higher-order co-moments, particularly co-kurtosis, so as to control joint tail risk and potentially allow for more aggressive leverage under stringent extreme-event constraints. Integrating a co-kurtosis term into the objective would enable the optimizer to penalize large simultaneous deviations more directly, thereby improving resilience in crises. Additional avenues include embedding context-dependent eigenvector adjustments to enhance signal extraction across regimes, coupling module parameters for joint adaptation to market states, unifying variance-based risk control with return forecasting via differentiable optimization layers, and explicitly incorporating market-impact models to capture liquidity costs under aggressive trading. Pursuing these directions promises to further strengthen the robustness, tail-risk management, and practical applicability of neural network–based portfolio optimization models.

\section*{Code Availability}
The source code implementing the proposed architecture is publicly available at 
\url{https://github.com/bongiornoc/Compact-RIEnet}. 
In addition, upon request, we can provide the calibrated neural network models used in this study to facilitate replication and further research.

\section*{Acknowledgments}
This work was performed using HPC resources from the ``Mésocentre'' computing center of CentraleSupélec and École Normale Supérieure Paris-Saclay supported by CNRS and Région Île-de-France (\url{http://mesocentre.centralesupelec.fr/}).
E.M.~acknowledges a fellowship funded by PNRR  for the PhD  DOT1608375  in Sistemi complessi per le scienze fisiche, socio-economiche e della vita of Catania University.





\end{document}